\documentclass[11pt]{article}

\usepackage{mytex}
\usepackage{eepic}

\begin{document}

\begin{center}
 {\bf Data Assimilation and Sensitivity  \\ of the Black Sea Model to Parameters. \\}
   \vspace{5mm}
  {\large E. Kazantsev }\\
 \vspace{5mm}
 {\it\small INRIA, MOISE team, 
 Laboratoire Jean Kuntzmann, Grenoble} \\
 \end{center} 

{\small
An adjoint based technique is applied to a Shallow Water Model in order to estimate influence of the model's parameters on  the solution. Among parameters  the bottom topography, initial conditions,   boundary conditions on rigid boundaries, viscosity coefficients  and the amplitude of the wind stress tension are considered.  Their influence is analyzed from different points of view.

Two configurations have been analyzed: an academic case of the model in a square box and a more realistic case simulating   Black Sea currents. It is shown in both experiments that  the boundary conditions near a rigid  boundary influence the most the solution. This fact points out the  necessity to identify optimal boundary approximation during a model development.}


\section{Introduction}

Thirty years ago model and data were considered as independent one on another. Observational data were interpolated on the model grid in order to provide the model with the initial conditions,  forcings and all the other necessary parameters. However, since the    pioneering work  \cite{Lor63} of Edward Lorenz, we know   that a geophysical fluid is extremely sensitive to initial conditions.  A  perturbation of  initial state may grow exponentially in time limiting the validity of the forecast. This discovery leads to understanding that observational data can not be considered as independent of the model.  We must perform a joint analysis of the  model and data in order to choose the optimal initial point for the model. 

This become possible by using variational  data assimilation technique,  first proposed in \cite{Ledimet82}, \cite{ldt86}, which is  based on the optimal control methods \cite{Lions68} and perturbations theory   \cite{Marchuk75}. This technique allows us to retrieve an optimal data for a given model from heterogeneous observational fields ensuring a better forecast.  

However, even now, all other forcings and parameters of the model are obtained from data by more or less sophisticated interpolation and they can not be considered as optimal for a given model. In the same time, we may suppose, that their influence on the models solution is as strong as the influence of initial state.  In this case, we should also analyze the possibility and utility to apply the data assimilation techniques to identify optimal values for all these parameters in order to improve the forecast. 

The purpose of this paper is to analyze the sensitivity of a Shallow-Water model and, in  particular, compare the influence of initial conditions with the influence of other parameters. Among these  parameters, we consider the boundary conditions on the rigid boundaries, bottom topography, empirical coefficients like reduced gravity, forcing amplitude and dissipation.  

\section{Sensitivity and control of the boundary conditions}

We shall focus our attention on the boundary conditions because (as we shall see later) they represent the most  unusual control variable. 

However, as it has been noted in  \cite{Leredde}, particular attention must be paid to the discretization process which must respect several rules because it is the discretization of the model's operators that takes into account the set of boundary conditions and introduces them into the model. Consequently,  instead of controlling boundary conditions them-self, it may be more useful to  identify optimal discretization of differential operators in points adjacent to  boundaries because this is  more general case. Indeed, boundary conditions participate  in discretized operators, but considering the discretization itself, we take into account additional parameters like the position of the boundary,  lack of resolution of the grid, etc. 

Boundary conditions are usually introduced into the model by a particular discretization of operators near the boundary. For example, taking into account the condition
$ u_0=0$ we can calculate the derivative at the point $x=h/2$ as $ \left. \der{u}{x}\right|_{1/2}=\fr{u_1}{h}$. 

In this paper, we  shall write the approximation of the derivative in a general form 
$$\left.\der{u}{x}\right|_{1/2}=\fr{\alpha_0+\alpha_1 u_1}{h}$$
Coefficients  $\alpha_0$ and $\alpha_1$ will be used as controls. That means we shall let them vary in the data assimilation procedure in order to find an optimal pair that realizes the minimum of the cost function.

\subsection{Example: one-dimensional wave equation}

In order to understand what happens when the data are assimilated to control the boundary conditions, we propose to  take a look on a scholar example: one-dimensional wave equation
written for $u=u(x,t)$ and $p=p(x,t)$ in the following way:
\beqr
\der{u}{t} &-& \der{p}{x}=0 \nonumber \\
\der{p}{t} &-& \der{u}{x}=0 \label{wave} 
\eeqr
This equation is defined on the interval $0<x<1$ with boundary conditions prescribed for $u$ only:
\beq u(0,t)=u(1,t)=0 \label{bc}\eeq

Initial conditions are prescribed for both $u$ and $p$
\beq u(x,0)=\bar u, \; p(x,0) = \bar p \label{ic} \eeq

The equation is discretized on a regular grid that is somewhat similar to Arakawa C grid \cite{ArakawaC} in two dimensions:

\begin{figure}[h]
\setlength{\unitlength}{0.7mm}
\newcount\indi
\newcount\num
\tiny
\begin{picture}(150,20)
\Thicklines
\put(5,10){\line(1,0){140}}
\put(5,8){\line(0,1){4}}
\put(145,8){\line(0,1){4}}
\thicklines

\multiput(5,9)(15,0){4}{\line(0,1){2}}
\indi=-1
\multiput(2,6)(15,0){4}{ 
\global\advance\indi by 1 $u_{\the\indi}$ }

\multiput(130,9)(-15,0){3}{\line(0,1){2}}
\put(144,6){ $u_{N}$ }
\indi=0
\multiput(127,6)(-15,0){3}{ 
\global\advance\indi by 1 $u_{N-\the\indi}$ }

\multiput(12,9)(15,0){3}{\line(1,1){2}}
\multiput(14,9)(15,0){3}{\line(-1,1){2}}
\indi=-1
\multiput(10,13)(15,0){3}{ 
\global\advance\indi by 1 \num=\indi \multiply\num by 2 \global\advance\num by 1 $p_{\the\num/2}$ }

\multiput(137,9)(-15,0){3}{\line(1,1){2}}
\multiput(139,9)(-15,0){3}{\line(-1,1){2}}
\indi=-1
\multiput(132,13)(-15,0){3}{ 
\global\advance\indi by 1 \num=\indi \multiply\num by 2 \global\advance\num by 1 $p_{N-\the\num/2}$ }
\end{picture} 

\end{figure}

Discrete derivatives  of $u$ and  $p$ are defined as follows 
\beq
\biggl(\der{p}{x}\biggr)_i =\fr{p_{i+1/2}-p_{i-1/2}}{h},\;\biggl(\der{u}{x}\biggr)_{i+1/2} =\fr{u_{i+1}-u_{i}}{h}
\nonumber
\eeq
in all internal points i.e. $2\leq i \leq N-2$ for $\biggl(\der{p}{x}\biggr)_i$ and $1\leq i \leq N-2$ for $\biggl(\der{u}{x}\biggr)_{i+1/2}$. 
Near the boundary, at points $1/2,1,N-1,N-1/2$, we   write the derivatives  in a general form, like  
\beq
\biggl(\der{p}{x}\biggr)_1= \fr{\alpha^p_0 p_{1/2}+ \alpha^p_1 p_{3/2}}{h}, \hspace{1cm}
\biggl(\der{u}{x}\biggr)_{1/2}=\fr{\alpha^u_0 + \alpha^u_1 u_{1}}{h} \label{bndsch1}
\eeq
considering $\alpha_0$ and $\alpha_1$ as the control coefficients. Leap-frog scheme was used for time stepping.

We introduce the simplest cost function that represents the distance between the model solution and observation at time $t$:
\beq
\costfun(\alpha)=\int\limits_0^T \int\limits_0^1 u(\alpha,x,t)-u^{obs}(x,t))^2+(p(\alpha,x,t)-p^{obs}(x,t))^2 dx dt \label{cf1d}
\eeq
and we calculate its gradient using the adjoint to the derivative of the solution with respect to control coefficients $\alpha^p,\alpha^u$:
 \beq
 \nabla\costfun = 2\int\limits_0^T {\biggl(\der{u(t),p(t)}{\alpha}\biggr)^* \vc{u(\alpha,x,t)-u^{obs}(x,t)}{p(\alpha,x,t)-p^{obs}(x,t)}} dt \label{gcf1d}
\eeq

Once we prescribe the initial conditions for the equation $$u(x,0)=sin(k\pi x) \;\; p(x,0)=cos(k\pi x),$$
we can calculate its exact solution: 
$$ u_{exact}(x,t)=\sqrt{2}sin(k\pi t-\pi/4) sin(k\pi x),\;\; 
 p_{exact}(x,t)=-\sqrt{2}cos(k\pi t-\pi/4)cos(k\pi x).$$
The exact solution is used as artificial observational data in this example. We perform the minimization of the cost function \rf{cf1d}.  The  minimization procedure used here was developed by Jean Charles Gilbert and  Claude Lemarechal, INRIA \cite{lemarechal}.  The procedure uses the limited memory quasi-Newton method.

The difference between the models solution and the exact one is shown in \rfg{1d}.
\begin{figure}[htbp]
\begin{center}
\begin{minipage}[t]{.55\textwidth}
  \centering \includegraphics[angle=0,width=.98\textwidth]{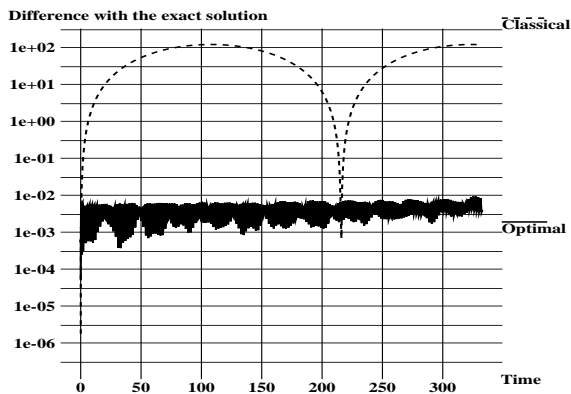} 
\end{minipage}
\end{center}
\refstepcounter{fig}
\label{1d}
\caption{Difference between the models solution and the exact one: classical BC --dashed line, optimal BC   -- solid line} 
\end{figure}

We see that optimal discretization of derivatives near the boundary brings the solution much closer to the exact solution, but the set of optimal coefficients $\alpha$ does not approximate a derivative:
\beq
\biggl(\der{u}{x}\biggr)_{1/2}=1.048\fr{u_1-u_0}{h}, \hspace{5mm}
\biggl(\der{p}{x}\biggr)_{1}=\fr{3.014 p_{3/2}-2.828 p_{1/2}}{h} \nonumber
\eeq

Neither expression for $\der{u}{x}$, nor for $\der{p}{x}$ has any reasonable order of approximation. The first one is of 0 order, the second is of -1 order. Moreover, while we get always the same formula for  $\der{u}{x}$,  approximation of the derivative of $p$ varies in different assimilation experiments. Assimilations performed with different assimilation windows, for example, result in different coefficients for $ \der{p}{x}$. In fact, any combination $\alpha^p_0\; , \alpha^p_1$ in \rf{bndsch1} may be found as the result of assimilation under condition 
\beq\alpha^p_1=-1.104 \alpha^p_0 -0.107.\label{p-line}\eeq
 This linear relationship has been obtained experimentally performing assimilations with all assimilation windows in range from 600 to 2400  time steps (with the  time step equal to $1/120$ of the  time unit).
 
To explain this strange result, we analyze the numerical solution of the equation. It is well known, the principal numerical error of the scheme is a wrong wave velocity. The wave speed, that must be equal to 1, is replaced by $\beta=\fr{h \sin(k\tau)}{2\tau \sin(kh/2)}$ which depends on the time step $\tau$ and the grid step $h$. For the given parameters ($k=3,\;  h=\fr{1}{30}$ and $\tau=\fr{1}{120}$), error in the wave velocity is equal to $3.09\tm 10^{-3}$

The data assimilation and control of the boundary derivatives can not modify numerical wave velocity. The only way for this control to get a better solution consists in modifying  the length of the interval. A numerical wave with wrong velocity will propagate on the interval with wrong length. But the length of the interval is adapted by data assimilation in order  to ensure the  wave with  numerical velocity propagates the modified interval in the same  time that the exact wave propagates the exact interval.  So far,  the control can not correct the error in the wave velocity, it commits another error in length in order to compensate  the first one as it is illustrated in \rfg{shiftgrid}.

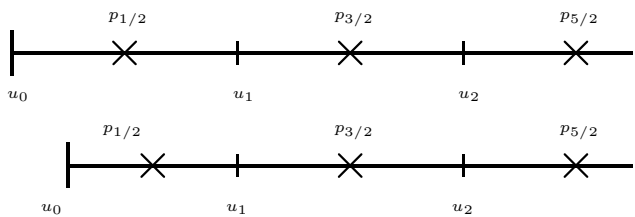
\begin{figure}[h]
\setlength{\unitlength}{1.5mm}
\newcount\indi
\newcount\num
\tiny
\begin{picture}(60,20)
\Thicklines
\put(5,15){\line(1,0){55}}
\put(5,13){\line(0,1){4}}
\thicklines
\multiput(5,14)(20,0){3}{\line(0,1){2}}
\indi=-1
\multiput(4,11)(20,0){3}{ 
\global\advance\indi by 1 $u_{\the\indi}$ }

\multiput(14,14)(20,0){3}{\line(1,1){2}}
\multiput(16,14)(20,0){3}{\line(-1,1){2}}
\indi=-1
\multiput(13,18)(20,0){3}{ 
\global\advance\indi by 1 \num=\indi \multiply\num by 2 \global\advance\num by 1 $p_{\the\num/2}$ }

\Thicklines
\put(10,5){\line(1,0){50}}
\put(10,3){\line(0,1){4}}
\thicklines
\put(10,4){\line(0,1){2}}
\put(25,4){\line(0,1){2}}
\put(45,4){\line(0,1){2}}
\indi=0
\put(7,1){ $u_{0}$ }
\multiput(24,1)(20,0){2}{\global\advance\indi by 1 $u_{\the\indi}$ }

\put(16.5,4){\line(1,1){2}}
\put(18.5,4){\line(-1,1){2}}
\multiput(34,4)(20,0){2}{\line(1,1){2}}
\multiput(36,4)(20,0){2}{\line(-1,1){2}}
\indi=0
\put(12.5,8){ $p_{1/2}$ }
\multiput(33,8)(20,0){2}{ 
\global\advance\indi by 1 \num=\indi \multiply\num by 2 \global\advance\num by 1 $p_{\the\num/2}$ }

\end{picture} 
\refstepcounter{fig}
\label{shiftgrid}
\caption{Modification of the intervals length.}
\end{figure}

Non uniqueness of optimal $\alpha^p_1$ and  $\alpha^p_0$  can be explained if we take into account that $p$ has also a form of cosine of $3\pi x$. Hence, at any time $p_{1/2}=A(t)\cos(3\pi h/2)$ and $ p_{3/2}=A(t)\cos(9\pi h/2)$ with some $A$ depending on time.  Their linear combination $\alpha^p_1 p_{1/2}+ \alpha^p_0 p_{3/2}$ can vanish if \beq\alpha^p_1=-\fr{\alpha^p_0}{4\cos^2(k\pi h/2)-3}.\label{zerocomb}\eeq
 Consequently, all couples  $\alpha^p_1\;, \alpha^p_0$ belonging to the  line that passes by the point 
$\alpha^p_0=-1.023\; , \alpha^p_1=1.023$ with tangent $-\fr{1}{4\cos^2(3\pi h/2)-3}= -1.108$
produce the same derivative. This line  coincides withing accuracy of computation  with the set \rf{p-line} obtained numerically.  Any point on this line gives coefficients $\alpha_p$ that theoretically  provide the same value of the derivative and the same value of the cost function.  

Of course, in this simple example we can avoid the ambiguity in the solution: it is sufficient to control only one coefficient $\alpha^p$ rather than two. But, in more complex problems, it may be difficult to locate and avoid the presence of kernels.

Consequently we can say that the data assimilation allows to place the boundary in the  optimal position resulting in a solution closer to the exact one.  Boundary control allows to compensate numerical errors committed in the interior of the domain, but it may be  difficult to understand the physical meaning of optimal coefficients $\alpha$ and
non-null kernels may exist leading to  non unique result.

More details of this study can be found in \cite{assimbc1}

\subsection{Shallow Water Model}

In this paper we consider   a  shallow-water model   written in a conservative form:
\beqr
\der{hu}{t}&=&  f hv - \der{}{x}\biggl( hu^2+gh(h-H)-\mu h\der{u}{x}\biggr)- 
 \nonumber \\&-&
  \der{}{y}\biggl(h u v-\mu h\der{u}{y}\biggr) 
   -\sigma hu+\tau_0 \tau_x, 
\nonumber \\
\der{hv}{t}&=&-f hu - \der{}{x}\biggl( h u v-\mu h\der{v}{x}\biggr)+
 \nonumber \\&-&
  \der{}{y}\biggl(\ hv^2+gh(h-H)-\mu h\der{v}{y}\biggr) 
    -\sigma hv  +\tau_0 \tau_y, 
\label{sw}  \\
\der{h}{t} &=& -\der{h u}{x}-\der{h v}{y}. \nonumber
\eeqr
where $hu(x,y,t)$ and $hv(x,y,t)$ are  two flux components that represent the product of the velocity by the ocean depth, $h(x,y,t)$, that corresponds to the distance from the sea surface to the bottom of the ocean. The sea  surface elevation is represented by the difference $h(x,y,t)-H(x,y)$, where $H(x,y)$ is the bottom topography.  The model is driven by the surface wind stress with components $\tau_x(x,y,t)$ and $\tau_y(x,y,t)$ normalized by $\tau_0$ and subjected to the  bottom drag that is parameterized by  linear terms $\sigma hu$ and $\sigma hv$. Horizontal eddy diffusion is represented by   harmonic operators  $div( \mu h \nabla u)$ and $div( \mu h \nabla v)$. Coriolis parameter is represented by the variable $f(y)$ that is equal to $f_0+\beta y$ assuming $\beta$-plane approximation. Parameter   $g$ is the reduced gravity.  The system is defined in some domain $\Omega$ with characteristic size $L$  requiring that   both  $hu$ and $hv$ vanish on the whole boundary of $\Omega$. No boundary conditions is prescribed for $h$.  Initial conditions are defined for all variables: $hu,\; hv$ and $h$. 

As usual, initial conditions are considered as the control parameter of the model in this paper. We study the sensitivity of the model to its initial point and assimilate data to find its optimal value. However, in addition to initial conditions, 
all other parameters of the model, and namely the discretization of operators near the boundary, its bottom topography $H(x,y)$,  scalar coefficients $\mu, \sigma, g$ and $\tau_0$,  are also considered as control variables. All of them are allowed to vary in the data assimilation procedure in order to bring them to their optimal values.

We discretize all variables of this equation on the regular  Arakawa's C-grid \cite{AL77} with constant grid step $\delta x=\fr{L}{N}$ in both $x$ and $y$ directions. 
Discretizing the system \rf{sw}, we replace the derivatives by their finite difference representations  $D_x$ and $D_y$ and   introduce  two    interpolations in $x$ and $y$ coordinates $S_x$ and $S_y$. Interpolations are necessary on the staggered grid  to calculate the variable's values in nodes where other variables are defined.  The discretized system \rf{sw} writes

\beqr
\der{hu}{t}&-& f S_xS_yhv+
D_x\biggl(\fr{(S_x hu)^2}{h}+gh(h-H(x,y))-\mu h D_x\fr{hu}{S_xh}\biggr)+ 
 \nonumber \\&+&
  D_y\biggl(\fr{(S_y hu\; S_x hv)}{S_x S_y h}-\mu (S_x S_y h)D_y\fr{hu}{S_x h}\biggr) 
  =  -\sigma hu+\tau_0\tau_x, 
\nonumber \\
\der{hv}{t}&+& f S_yS_x hu +D_x\biggl(\fr{(S_y hu\; S_x hv)}{S_x S_y h}-\mu (S_xS_y h) D_x\fr{hv}{S_y h}\biggr)+ 
 \nonumber \\&+&
  D_y\biggl(\fr{(S_y hv)^2}{h}+gh(h-H(x,y))-\mu h D_y\fr{hv}{S_y h}\biggr) 
  =  -\sigma hv  +\tau_0\tau_y, 
\label{sw-grid}  \\
\der{h}{t} &=&- D_x hu - D_y hv. \nonumber
\eeqr
Discretized operators $D_x, D_y$ and $S_x, S_y $ are defined in a classical way at all internal points of the domain. For example,  the second order derivative and the interpolation operator of the variable $hu$ defined at corresponding  points  write
\beqr
(D_x hu)_{i-1/2,j-1/2}&=&\fr{hu_{i,j-1/2} -hu_{i-1,j-1/2} }{\delta x}  \mbox{ for } i=2,\ldots , N-1,
 \nonumber \\
(S_x hu)_{i-1/2,j-1/2}&=&\fr{hu_{i,j-1/2} +hu_{i-1,j-1/2} }{2}\mbox{ for } i=2,\ldots , N-1.  \label{intrnlsch} 
\eeqr

   Discretization of operators in the  directly adjacent to the boundary nodes are  different from \rf{intrnlsch} and represent the control variables in this study. In order to obtain their  optimal values  assimilating external data, we suppose nothing about derivatives and interpolations near the boundary  and  write them in a general form
\beqr
(D_x hu)_{1/2,j-1/2}&=&\alpha_{0}^{D_x^{hu}}+\fr{\alpha^{D_x^{hu}}_{1} hu_{0,j-1/2} +\alpha^{D_x^{hu}}_{2} hu_{1,j-1/2}}{\delta x} \nonumber\\
(S_x hu)_{1/2,j-1/2}&=&\alpha_{0}^{S_x^{hu}}+\fr{\alpha^{S_x^{hu}}_{1} hu_{0,j-1/2} +\alpha^{S_x^{hu}}_{2} hu_{1,j-1/2}}{2}
 \label{bndschsw}
\eeqr

This formula represents a linear combination of values of $hu$ at two  points adjacent to the boundary with  coefficients $\alpha$. The constant $\alpha_0$ may be added in some cases to simulate non-uniform boundary conditions like $hu(0,y)=\alpha_0\neq 0$.

We distinguish $\alpha$ for different variables and different operators allowing different controls of  derivatives   because of the different nature of these variables and different boundary conditions prescribed for them. It is obvious, for example, that the approximation of the derivative $D_x$ in the first equation may differ from the approximation of  $D_x$ in the third one. Although both operators represent a derivative,  boundary conditions for $hu$ and $h$ are different, these derivatives are defined at different points, at different distance from the boundary. Consequently, it is reasonable to let them be controlled separately and  to assume that their optimal approximation may be different with distinct coefficients $\alpha^{D_x^{hu}}$ and $\alpha^{D_x^{h}}$.

Time stepping of this model is performed by the leap-frog scheme. 
The first time step is splitted into two Runge-Kutta stages in order to ensure the second order approximation.

As well as before, the approximation of the derivative introduced by \rf{intrnlsch} and \rf{bndschsw} depends on variables $\alpha$. These variables  are added to the set of control variables enumerated above. Operators  are allowed to change their properties near boundaries in order to find the best fit with requirements of the model and data.   To assign all control variables   we shall perform data assimilation procedure and find their optimal values.  Variational data assimilation is usually performed by minimization of the specially introduced cost function. The minimization is achieved using the gradient of the cost function that is usually determined by the run of the adjoint to the tangent linear model.

 To define the cost function we introduce dimensionless  state vector $\phi$ that is composed of three variables of the model $\phi=\{w_{hu}hu,w_{hv}hv,w_h h\}^t$ weighted by coefficients $w$. These weights  are used to normalize values of the flux components by  $w_{hu}=w_{hv}=\fr{1}{H_0\sqrt{gH_0}}$ and the Sea surface elevation by $w_h=\fr{1}{H_0} $. The distance between the model solution and observations is defined as the Euclidean norm of the difference

\beqr
\xi^2&=&\xi^2(\phi(p,t))=\sum_{k} (\phi_{k}- \phi^{obs}_{k})^2 =
\label{xiphi}\\
&=&w_{hu}^2\sum_{i,j} (hu_{i,j}- hu^{obs}_{i,j})^2 + w_{hv}^2\sum_{i,j}(hv_{i,j}- hv^{obs}_{i,j})^2+w_h^2\sum_{i,j} (h_{i,j}- h^{obs}_{i,j})^2. \label{xi}
\eeqr
In this expression, we emphasize implicit  dependence of $\xi$  on time and on the set of the control parameters $p$ that is composed of 
\begin{itemize}
\item the set of initial conditions of the model $ \phi_0=\{hu\mid_{t=0},\; hv\mid_{t=0},\; h\mid_{t=0}\}$,
\item  the set of the coefficients $\alpha$ that controls the discretizations of operators near the boundary,
\item the bottom topography $H(x,y)$
\item four scalar parameters $\sigma, \mu, g, \tau_0$.
\end{itemize}
Taking into account the results obtained in \cite{sw-lin}, we define the cost function as 
\beq
\costfun(p)=\int\limits_0^T  t \xi^2(\phi(p,t)) dt \label{costfn}
\eeq
that gives  higher  importance to the difference $\xi^2$ at the end of assimilation interval.  

It should be noted here, that this cost function can  only  be used in the case of  assimilation of a perfect artificially generated data. When we assimilate some kind of real data that contains errors of measurements and is defined on a different grid, we should add some regularization term to the cost function (like the distance from the initial guess) and use some more appropriate norm instead of the Euclidean one (see, for example  \cite{UnifNotat} for details).  

The $n$th component of the gradient of the cost function can be calculated as the Gateaux derivative of an implicit function:
 \beqr
 (\nabla\costfun)_n&=&\der{\costfun}{p_n}=\int\limits_0^T t\biggl(  \der{\xi^2}{p_n}\biggr) dt=\int\limits_0^T t\biggl(  \sum_k\der{\xi^2}{\phi_k}\der{\phi_k}{p_n}\biggr) dt =\nonumber \\
 &=&2\int\limits_0^T t\biggl(  \sum_k (\phi_k-\phi^{obs}_k)  \der{\phi_k}{p_n}\biggr) dt\label{nabj}
\eeqr 
because the  derivative $\der{\xi^2}{\phi_k}$ can easily be calculated from \rf{xiphi}: $\der{\xi^2}{\phi_k}=2 (\phi_k-\phi^{obs}_k) $. The second term in \rf{nabj},  $\der{\phi_k}{p_n}$, represents the matrix of the tangent linear model that relates the perturbation of the parameter $p_n$ and the perturbation of $k$th component of the model state vector $\phi_k$. This relationship, of course, is assumed in the linear approach, that means it is only valid for infinitesimal perturbations. 

 In the classical case, when initial conditions are considered as the  only control variable, the derivative  $\der{\phi(t)}{p}=\der{\phi(t)}{\phi_0}$  is the classical tangent model that describes the temporal evolution of a small error in the initial model state. The matrix is a square matrix that is widely studied in numerous sensitivity analyses. Its singular values at infinite time limit are related to well known Lyapunov exponents that determine the model behavior (chaotic or regular) and the dimension of it's attractor.
 
 In our case,  the matrix $\der{\phi(t)}{p}$ is rectangular. It describes the evolution of an infinitesimal error in any parameter (including initial state). However, we can study it's properties   in the similar way as we do with the classical tangent linear model. Its structure and composition is described in \cite{sw-lin} for the case of using coefficients $\alpha$ as control parameters and in \cite{assimtopo} for the case when the bottom topography is used to control the model's solution. 
 
The product  $\sum_k  (\phi_k-\phi^{obs}_k) \der{\phi_k}{p_n}$ in \rf{nabj} represents an unusual  vector-matrix product. To calculate this product directly we would  have to evaluate all the elements of the matrix. This would require as many tangent model runs as the size of the state vector is. So, instead of the tangent model, we shall use the adjoint one that allows us to get the result by one run of the model. Backward in time adjoint model  integration that starts from $(\phi-\phi^{obs}) $  provides immediately the product  $\biggl( \der{\phi}{p}\biggr)^*(\phi-\phi^{obs}) $ which is exactly equal to  $  (\phi-\phi^{obs}) \der{\phi}{p}$ in \rf{nabj}.  

  Using these notations, we write
\beq
\nabla\costfun= 2 \int\limits_0^T t \biggl( \der{\phi(t)}{p}\biggr)^*(\phi( p,t)-\phi^{obs}(t)) dt  \label{grad}
\eeq
where the expression in the integral is the result of the adjoint model run from $t$ to 0 starting from the vector $ (\phi( p,t)-\phi^{obs}(t))  $.

Tangent and adjoint models have been automatically generated by the Tapenade software \cite{Hascoet04},\cite{Tber07} developed by the TROPICS team in INRIA. This software analyzes the source code of the nonlinear model and produces codes of it's derivative $ \der{\phi}{p}$ and of the adjoint $\biggl( \der{\phi}{p}\biggr)^*$. 

This gradient is used in the minimization procedure that is implemented in order  to find the minimum of the cost function:
\beq
\costfun(\bar p) = \min\limits_p \costfun(p) \label{min}
\eeq
Coefficients $\bar p$  are considered as coefficients achieving an  optimal parameters for the model. 
As it has been already noted, the set of parameters $p$ is composed of   the set of initial conditions of the model $ \phi_0$,
  the set of the coefficients $\alpha$ that controls the discretization of operators near the boundary,
the bottom topography $H(x,y)$ and 
 four scalar parameters $\sigma, \mu, g, \tau_0$. We shall minimize the cost function  controlling either the  total set of available parameters $p$ or any possible subset, comparing the efficiency of the minimization.

We use the  minimization procedure  developed by Jean Charles Gilbert and  Claude Lemarechal, INRIA \cite{lemarechal}.  The procedure uses the limited memory quasi-Newton method.

In addition to the data assimilation, we perform also the sensitivity study of the model solution to parameters enumerated above. We are looking for a perturbation in the model's parameters $\delta p$  that, for a given small  norm,  maximizes the norm of the perturbation of the solution at time $t$.  
\beq
\lambda(t)=\max\limits_{\delta p} \fr{\norme{\delta\phi(t)}}{\norme{\delta p}}
\eeq
We can note that we already have  all the necessary software to estimate $\lambda(t)$. Tangent linear model $\biggl( \der{\phi(t)}{p}\biggr)$ allows us to calculate $\delta\phi(t)= \biggl( \der{\phi(t)}{p}\biggr)\delta p$. Using the scalar product that corresponds to the  norm  in the definition of the distance $\xi$ \rf{xi}, we can write 
\beqr
\lambda(t)&=&\max 
\fr{\spm{\delta\phi(t)}{\delta\phi(t)}}{\spm{\delta p}{\delta p}}= 
\max \fr{\spm{\biggl( \der{\phi(t)}{p}\biggr)\delta p}{\biggl( \der{\phi(t)}{p}\biggr)\delta p}}{\spm{\delta p}{\delta p}}= \nonumber \\
&=&\max \fr{\spm{\biggl( \der{\phi(t)}{p}\biggr)^*\biggl( \der{\phi(t)}{p}\biggr)\delta p}{\delta p}}{\spm{\delta p}{\delta p}} \label{relritz}
\eeqr
This expression is a well known  Rayleigh-Ritz ratio which is equal to the largest eigenvalue of the problem
\beq
\biggl( \der{\phi(t)}{p}\biggr)^*\biggl( \der{\phi(t)}{p}\biggr) \vartheta= \lambda(t)\vartheta \label{eigval}
\eeq
So far, we need just the maximal eigenvalue and the matrix of the problem is a self-adjoint positive definite matrix, we can solve the problem \rf{eigval} by the power method performing successive iterations 
$$
\vartheta_{n+1}=\fr{ \biggl( \der{\phi(t)}{p}\biggr)^*\biggl( \der{\phi(t)}{p}\biggr) \vartheta_n}{\norme{ \biggl( \der{\phi(t)}{p}\biggr)^*\biggl( \der{\phi(t)}{p}\biggr) \vartheta_n}},\;\; \vartheta_0=\mbox{random vector} 
$$
In the limit,  the denominator of the right-hand-side tends to the largest eigenvalue and  $\vartheta_{n}$  --- to the corresponding  eigenvector of the matrix. 
The principal advantage of this method consists in the fact that we do not need to calculate the matrix itself, we just need  a matrix-vector  product. So far, we have both codes for the  tangent and adjoint models, we can successively run these models and get the left-hand side of \rf{eigval}. 

We should note here that when the initial conditions of the model are used as the control parameters (i.e. $\delta p=\delta\phi(0)$), the sensitivity characteristics $\lambda(t)$ are all close to one when $t \longrightarrow 0$. It is evident because the perturbation has no time to be transformed by the model's dynamics and we get $\delta\phi(t)\mid_{t \longrightarrow 0}=\delta\phi(0)=\delta p$. 

When any other model parameter is used as the control and the error growing time is small, all   $\lambda(t)$ are vanishing. This is also clear: the model's dynamics has no time to transmit the perturbations from the parameters to the solution. The perturbation of the solution remains, consequently, close to zero as well as the value of  $\lambda(t)\mid_{t \longrightarrow 0}=0$.  

In order to make   the behavior of the sensitivity characteristics  uniform with different parameters, we shall use  $\lambda(t)-1$ every time when the initial model's state is considered as the control parameter.  

\section{Configurations.}
\subsection{ Model in a square box. }
We start from the data assimilation in frames of the very well studied "academic" configuration. 
Several experiments have been performed with the  model  in a square box of side length $L=2000$ km driven by a steady, zonal  wind forcing with a classical sinusoidal profile
$$
\tau_x=\tau_0 \cos \fr{2\pi (y-L/2)}{L}
$$
that leads to the formation of a double gyre circulation \cite{LPV}. The attractor of the model and the bifurcation diagram  in a similar configuration has been described in \cite{simmonet2}. Following their results, we intentionally chose the model's parameters to ensure chaotic behavior. The maximal wind tension on the surface is taken to be $\tau_0=0.5\fr{dyne}{cm^2}$.  The coefficient of Eckman dissipation and the  lateral friction coefficient  are chosen as
$\sigma=5\tm 10^{-8}s^{-1}$ and  $\mu=200\fr{m^2}{s}$ respectively. 

As it has been already noted, the Coriolis parameter is a linear function in $y$ with  $f_0=7\tm 10^{-5}s^{-1}$ and $\beta=2\tm 10^{-11} (ms)^{-1}$. The reduced gravity and the depth are respectively equal to $g=0.02\fr{m}{s^2},\;H_0=1000m$.

The resolution of the model in this section is intentionally chosen to be too coarse to resolve the Munk layer \cite{Munk}  that is characterized by the local equilibrium between the $\beta$-effect and the lateral dissipation. Its characteristic width is determined by the Munk parameter $ d=2\biggl(\fr{\mu}{\beta}\biggr)^{1/3}$ which is equal to 42 km in the present case.
The model's grid is composed of 30 nodes in each direction, that means the grid-step is equal to 67 km, that is more than  the Munk parameter.  Thus, there is only one grid node in the layer and the solution exhibits spurious oscillations near the western boundary due to unresolved boundary layer.  

Artificial ``observational`` data are generated by
the same model with all the same parameters but with 9 times finer resolution  (7.6 km  grid step). The  fine resolution model, having 7 nodes in the Munk layer, resolves  explicitly the layer and must have no spurious oscillations. All nodes of the coarse grid belong to the fine grid, consequently, we do not need to interpolate ''observational'' data to the coarse grid. We just take values in  nodes of the high resolution grid that correspond to nodes on the coarse grid. 

The model on the fine grid has been spun up from the rest state during 3 years. The end of spin up was used as the initial state for the further integration of the model. From the result of this integration we have extracted values of all three variables at all grid points that belong to the coarse grid (as it has been noted, the grids have been chosen so, that all grid points of the coarse grid belong to the fine grid). This set   is used as artificial observations in the following experiments. 

So far the model is nonlinear with intrinsicly instable solution, there is no hope to obtain close solutions in  long time model runs because any difference (even infinitesimal) between two models grows exponentially in time. Consequently, we have to confine our study to the analysis of a short time evolution of the model's solution simulating the forecasting properties of the model.

  As the initial guess for the initial conditions we use the state vector of the high resolution model reduced on the coarse grid. This state is also used as the initial conditions in all other   assimilation experiments with other control parameters.  Noted above values of the model's parameters (flat bottom topography, linear in $y$ Coriolis parameter and  scalar parameters ($\mu, \; \sigma, \; \tau_0,\; g$) are used as the initial guess in the experiments that control these parameter, otherwise we simply use these parameters in the model.

\subsection{Model of the Black Sea. }

In this section we use the same model, but all the parameters are defined to describe the upper layers circulation of  the Black sea. Configuration of the model  and  observational data have been kindly provided by Gennady Korotaev from the  Marine Hydrophysical Institute, National Academy of Sciences of
Ukraine, Sevastopol, Ukraine. This configuration is described in \cite{korot-model}. 

The model grid counts $141\tm 88$ nodes  that corresponds to the  grid box of dimension 7860 m and 6950 m in $x$ an $y$ directions respectively.   15 minutes  time step is
used for integration of the model.  The Coriolis parameter is equal to  $f_0= 10^{-4}s^{-1}$ and $\beta=2\tm 10^{-11} (ms)^{-1}$. Horizontal viscosity is taken as $\mu=50 m^2 s^{-1}$. Using a typical density difference between upper and underlying layers of  $3.1 kg/m^3$, and unperturbed layer thickness of $H_0 = 150 m$, the Rossby radius of deformation is estimated
at about 22 km and the reduced gravity value $g=0.031 m/s^2$. The grid therefore resolves the mesoscale processes reasonably well.

The model has been forced by the ECMWF wind stress data, available as daily averages for the years 1988 through
1999. 
Dynamical sea level reconstructed in \cite{korot-altim} was used as  observational data in this section. These data   have been collected in ERS-1 and TOPEX/Poseidon missions and preprocessed by the NASA Ocean Altimeter Pathfinder Project, Goddard Space Flight Center. Observational data are available from the 1st May 1992 until 1999. These data have been linearly interpolated to the model grid.

So far  the sea surface elevation is the only observational variable available in this experiment, we put $w_{hu}=w_{hv}=0$ in \rf{xi}. Consequently, the difference between the model's solution and observations is calculated taking into account the variable $h$  only.  

As it has been already noted, absence  of  observational data for the velocity fields brings   us to modify the cost function. We have to add the background term in the cost function in order to  require the velocity field to be sufficiently smooth. Otherwise, lack of information about velocity components in observational data would result in a spuriously irregular fields obtained in assimilation. To ensure necessary regularity of $hu$ and $hv$ we add the distance from the initial guess to the cost function \rf{costfn}. In order to emphasize the requirement of smoothness,  this distance is measured as an enstrophy of the difference between the initial guess and current state: 
\beq
\costfun_{smooth}=  \sum_{i,j}\biggl(\der{(hv_{i,j}-hv_{i,j}^0)}{x}-\der{(hu_{i,j}-hu_{i,j}^0)}{y}\biggr)^2  \label{costfn-smooth}
\eeq
where $hu^0,\; hv^0$ denote  flux components of the initial guess of the minimization procedure. 

Moreover, using real observational data requires to add at least one another term to the cost function. One can see in 
the Figure 2 in \cite{korot-altim},  spatially averaged  sea surface elevation of the Black sea exhibits a well distinguished seasonal cycle. That means the mass is not constant during a year, it decreases in autumn and increases in spring. Consequently, if we assimilate data during a short time (a season or less), we assimilate also the information about the  mass flux specific for this season. This flux can not be corrected later by the model because the discretization of operators near the boundary (that controls the mass evolution) is obtained once for all seasons.  The mass variation of the Black sea  reaches 25 centimeters of the sea surface elevation. Assimilating data  within one season may, consequently,  result in a persisting increasing or decreasing of the seal level of order of 50 cm  per year.   To avoid this spurious change of the total mass, we must either take the assimilation window of at least one year, or prescribe the mass conservation to the model's scheme. One year assimilation window is computationally expensive and is not justified by the model's physics. On the other hand, prescribed mass conservation removes just the sinusoidal seasonal variation, allowing us to keep all other processes and to choose any assimilation window we need. 

To correct the mass flux of the model, we add the following term to the cost function
\beq
\costfun_{mass}= \int\limits_0^T \biggl(\sum_{i,j}(h_{i,j}(t)-h_{i,j}(0))\biggr)^2 dt \label{costfn-mass}
\eeq
Similarly to \rf{costfn-smooth}, this term also ensures the regularity of the solution. It can be noted here that  other terms may be added to the cost function in order to make a numerical scheme  energy and/or enstrophy conserving, but we do not use them in this paper. 

The total cost function in this section is composed of three parts: \rf{costfn}, \rf{costfn-smooth} and \rf{costfn-mass}:
\beq
\costfun_{total}= \costfun+\gamma_1\costfun_{smooth}+\gamma_2\costfun_{mass}\label{costfn-total}
\eeq
Coefficients $\gamma$ are introduced to weight the information that comes from observational data (with $\costfun$) and an a priori knowledge about mass conservation and regularity of the solution. 

This modification of the cost function results, of course, in additional terms in the gradient:
\beq
\nabla  \costfun_{total}= \nabla\costfun+ 2\gamma_1 \biggl( D_y^*D_y (hu-hu_0) + D_x^*D_x (hv-hv_0)\biggr) 
+2\gamma_2 \sum_{i,j} \biggl(\eta_{i,j}(t)-\eta_{i,j}(0)\biggr).
\eeq

The model is spun up  from the beginning of 1988 to May 1992 using the wind tension data on the surface. The state corresponding to the 1st of May 1992 12h GMT is used as the initial guess in the data assimilation procedure controlling initial conditions of the model.  The assimilation controls the initial conditions $\phi_0$ only with the assimilation window $T=1$ day and the regularization parameter $\gamma_1=0.04$.   Such a short window was chosen in order to get almost instantaneous state of the model to be used in further experiment as an initial state.

The behavior of the model  solution is not chaotic in this configuration. Variability of the model is generated directly by the variability of the wind stress on the surface.  Consequently, we can compare particular trajectories of the model on any time interval because their evolution is stable without exponential divergence. Thus, we can hope that  assimilating data in a relatively short window  allows us to bring the model's solution closer to observation for a long integration period. 

 The minimization of the cost function has been accompanied by the mass preserving correction \rf{costfn-mass} with $\gamma_2= 0.01$.

\section{Sensitivity analysis.}

The flexibility of the model is illustrated in \rfg{valj}. We perform the data assimilation experiment in two configurations  using parameters described above as initial guess.  Due to high CPU time of the data assimilation, we limit the number of iterations of the minimization procedure by 20. Thus, we have similar  and reasonable computational cost in each experiment.

In both configurations we examine the evolution of the distance "model--observations" $\xi(t)$ during assimilation and after the end of assimilation. Assimilation window has been chosen as 5 days in the square box configuration and $T=30$ days for the Black sea model. The distance is examined over longer intervals: 20 days in the first case and 1 year in the second one. 

\begin{figure}[h]
  \begin{center}
  \begin{minipage}[r]{0.48\textwidth} 
  \centerline{\includegraphics[angle=0,width=0.95\textwidth]{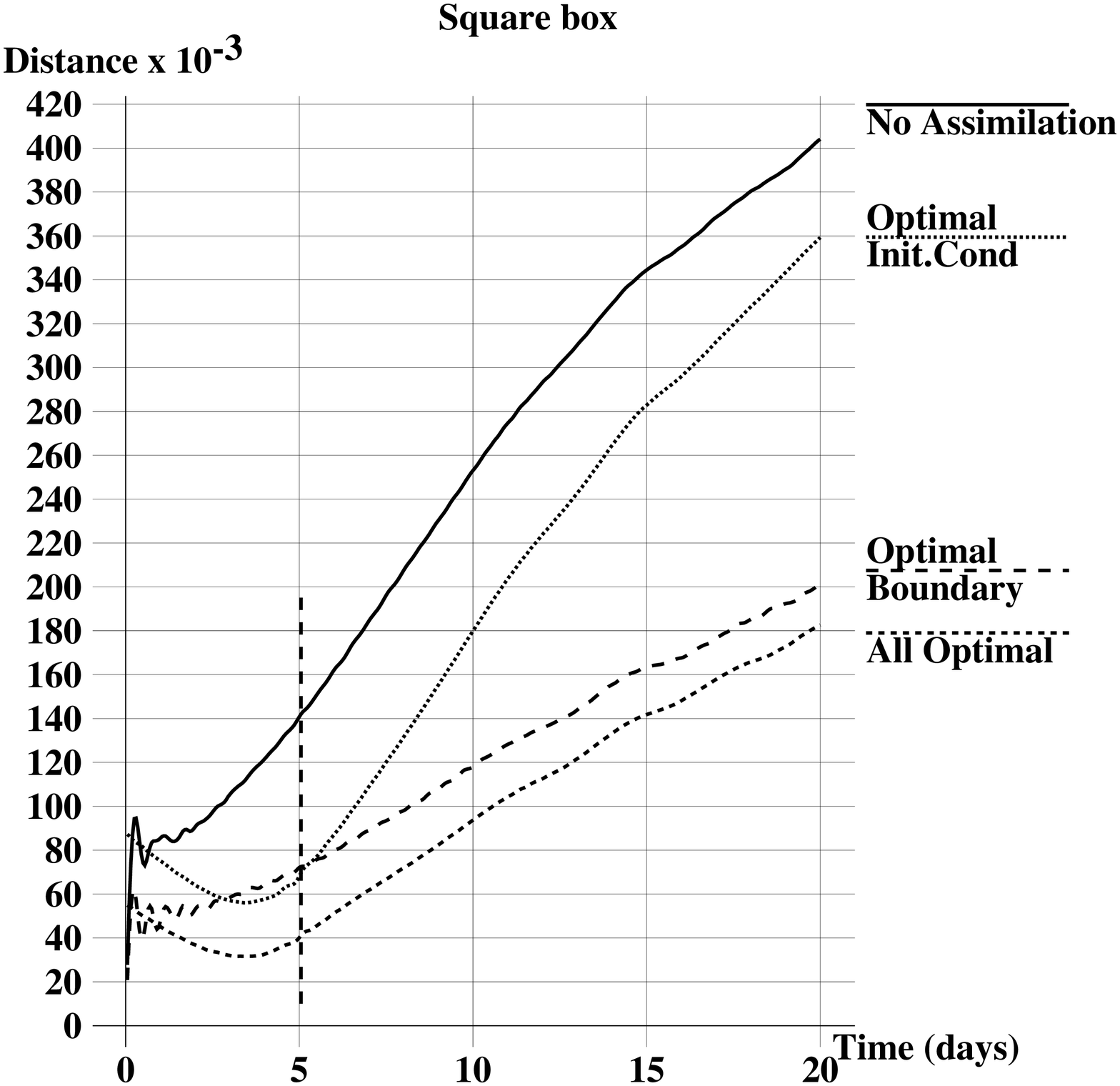}}
  \end{minipage} 
   \begin{minipage}[r]{0.48\textwidth} 
  \centerline{\includegraphics[angle=0,width=0.95\textwidth]{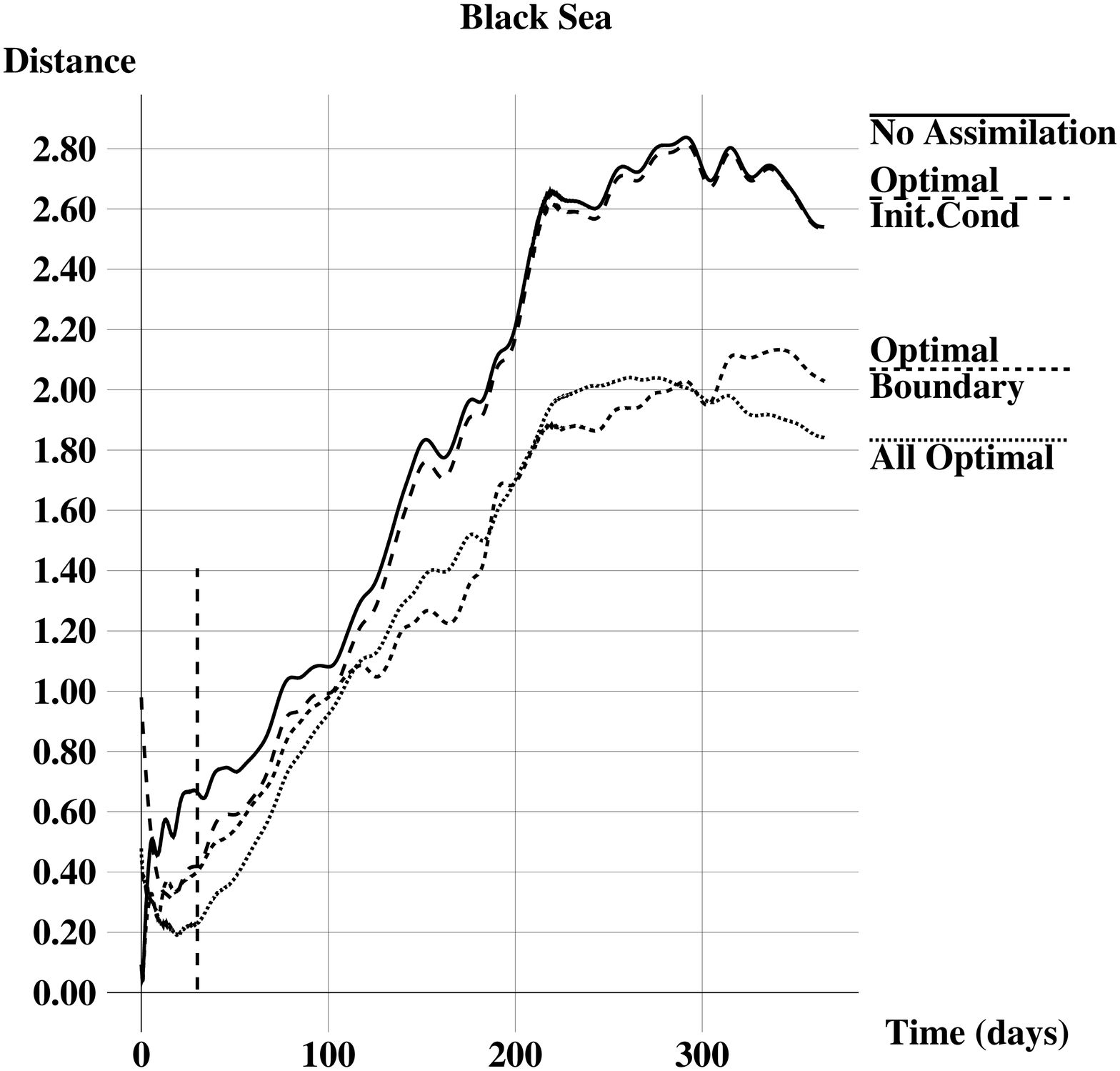}}
  \end{minipage} 
  \end{center} 
  \caption{Distance between the  model solution and observations for the model in the square box (left) and the model of the upper layer of Black Sea (right).   }  
\label{valj}
\end{figure}

Analyzing the difference between the  model solution and observations shown in \rfg{valj}, we see that in the assimilation window the model is almost equally flexible with respect to both  initial and boundary conditions. Data assimilation allows us to reduce the distance between the  model solution and observations at the end of the window approximately twice in both configurations. 
The only difference that can be seen in the assimilation window  is that non-optimal initial point leads to the spurious oscillations of the solution. These oscillations occur in both configurations and show us the necessity to identify the optimal initial point. 

However, the influence of  parameters is significantly different beyond the window. While the solution with optimal  initial point tends towards the solution obtained without any data assimilation,  optimal set of boundary conditions ensures a new solution that is much closer to observational data. That means the control of boundary conditions allows us to improve a long-range forecasting quality of the model.

The third way of the sensitivity analysis consists in solving of the eigenvalue problem \rf{eigval} and analyzing  $\lambda(t)$ on different  scales of error growing time from about $10$ minutes  ($10^{-3}$ day) to approximately one year. 
As it has been already noted, $\lambda(t)-1$ is plotted in the case when initial conditions are considered as the parameter.

\begin{figure}[htbp]
  \begin{center}
  \begin{minipage}[r]{0.45\textwidth} 
  \centerline{\includegraphics[angle=0,width=0.95\textwidth]{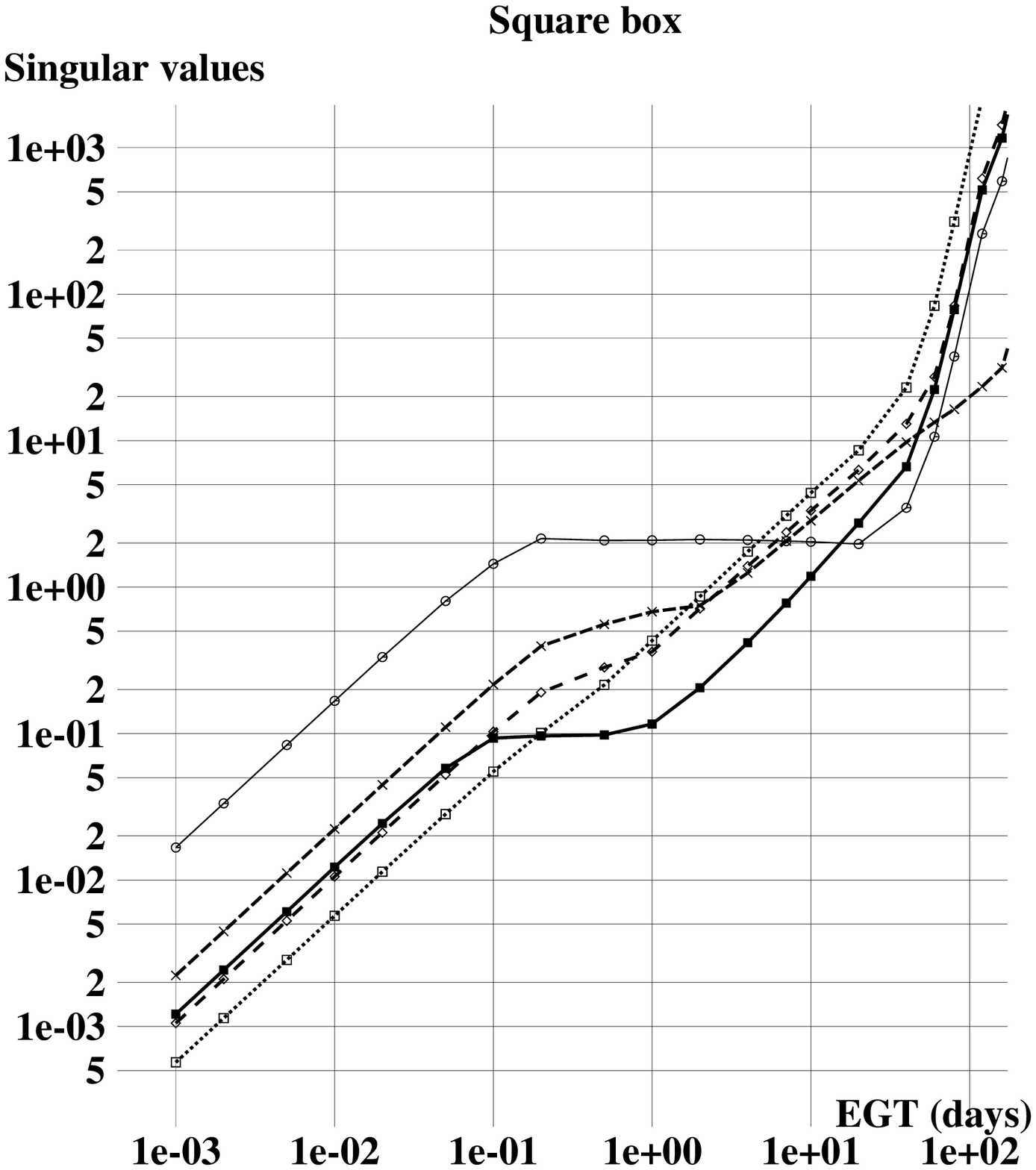}}
  \end{minipage} 
   \begin{minipage}[r]{0.45\textwidth} 
  \centerline{\includegraphics[angle=0,width=0.95\textwidth]{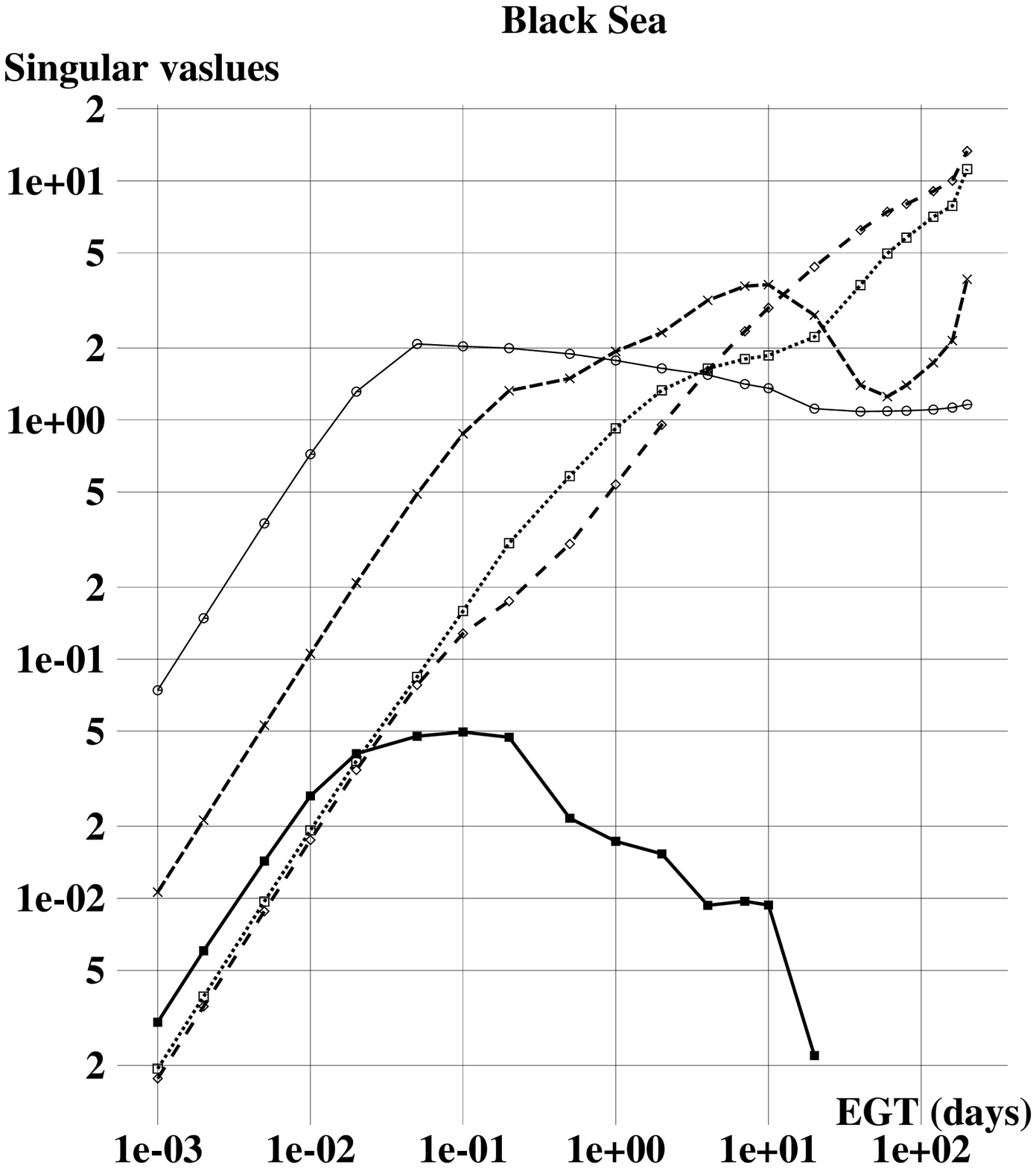}}
  \end{minipage} \\
  \begin{minipage}[r]{0.45\textwidth} 
  \centerline{\includegraphics[angle=0,width=0.95\textwidth]{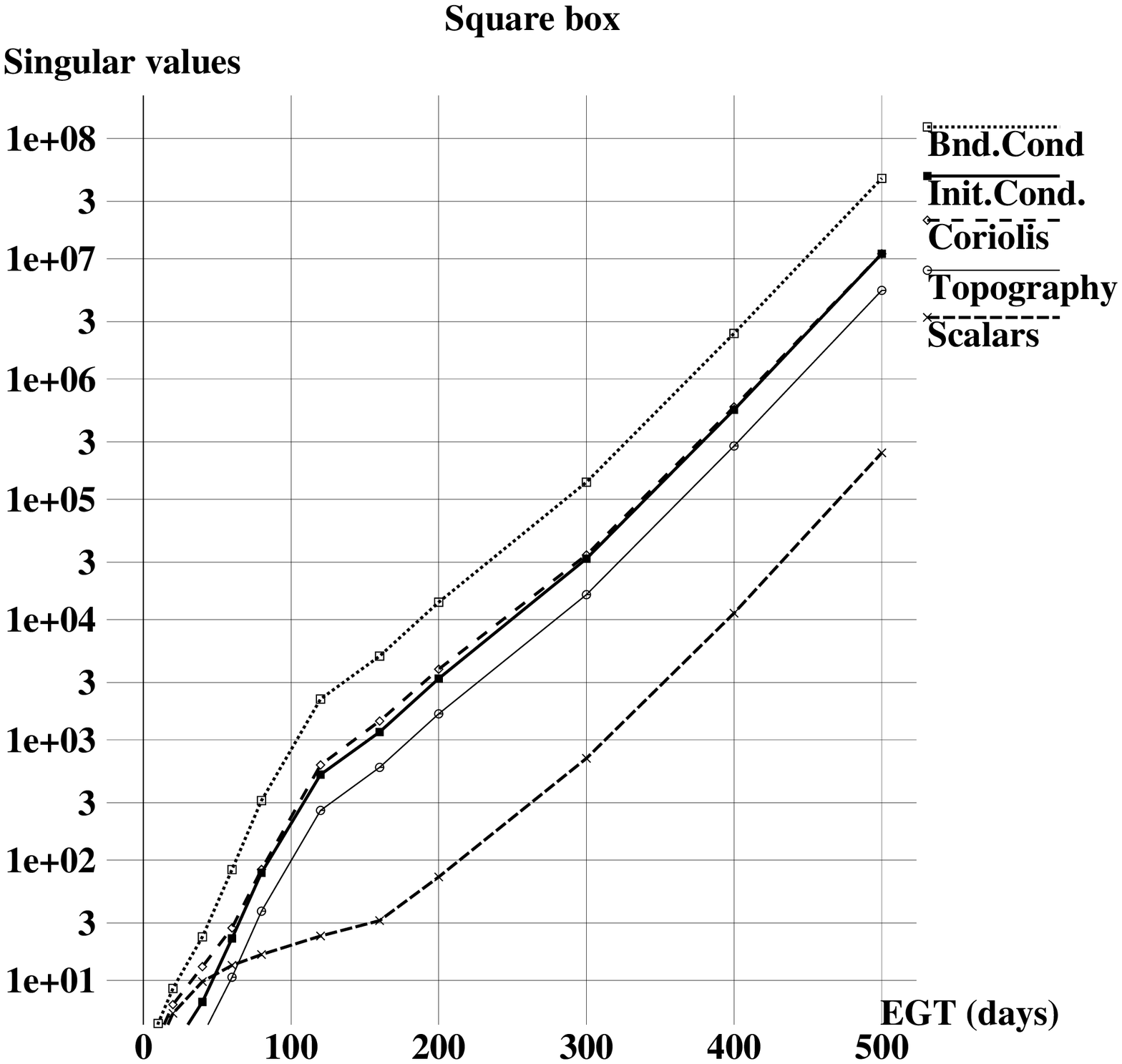}}
  \end{minipage} 
   \begin{minipage}[r]{0.45\textwidth} 
  \centerline{\includegraphics[angle=0,width=0.95\textwidth]{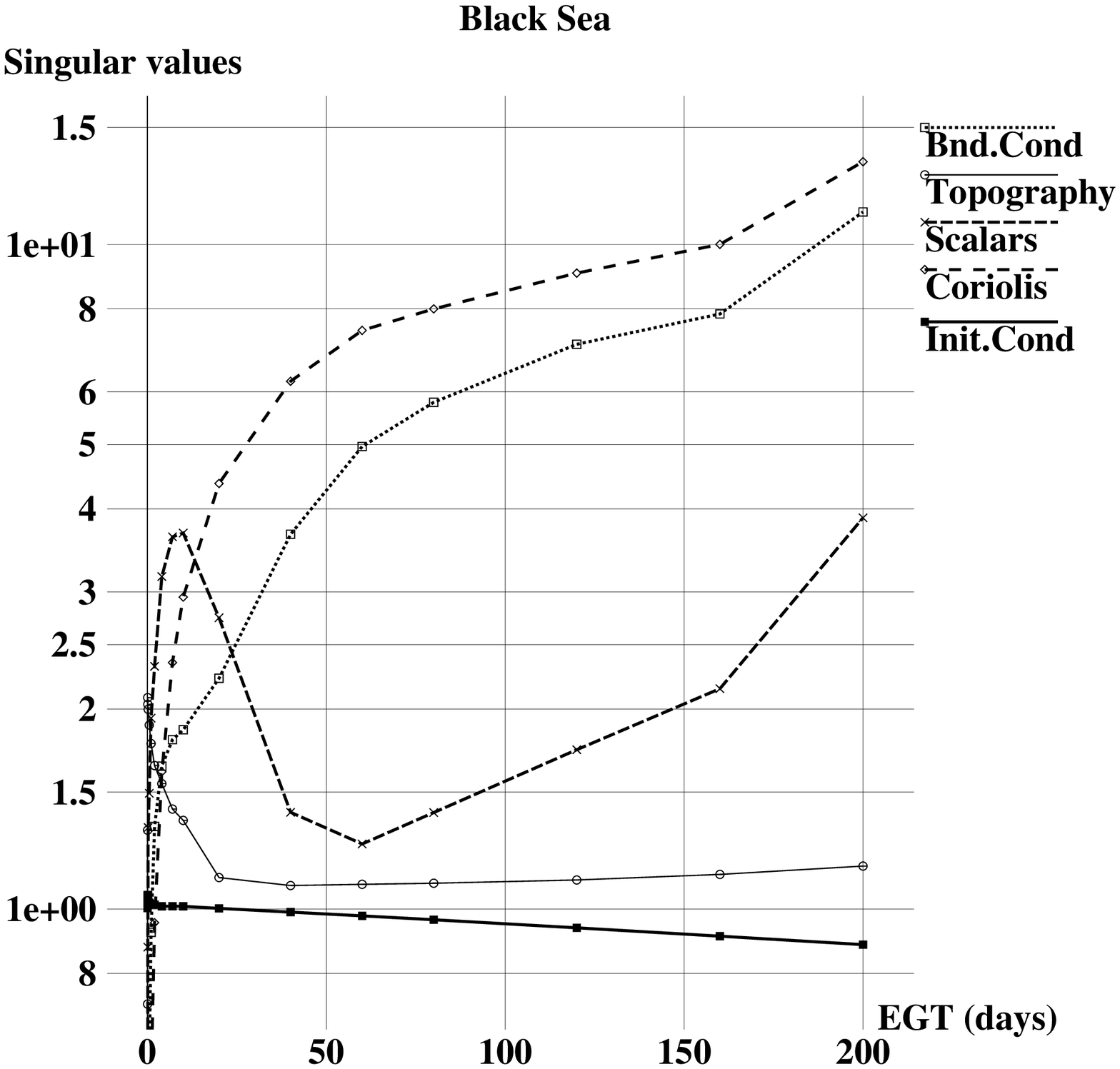}}
  \end{minipage} 
  \end{center} 
  \caption{Finite time Lyapunov exponents on short (above) and long (below) time scales: model in the square box (left) and the model of the upper layer of Black Sea (right).   }  
\label{evt}
\end{figure}

Analyzing the figure \rfg{evt}, we can see that three  time scales can be clearly distinguished  for the sensitivity characteristics of the model in both configurations. The first, short time scales, approximately from 0  up to  2-3 hours is characterized by the  linear growth of   $\lambda(t)$.  Indeed, the model behaves as a linear model on this scales,  the model's solution can be well approximated by just one step of the numerical time scheme. 

The second time scale  that can be distinguished in the  figure \rfg{evt} corresponds to error growing times from 2-3 hours to 10 days. On these time scales we see slower growth of the sensitivity characteristics $\lambda(t)$ and, sometimes, no growth at all. These time scales are characterized by the modification of the stable-instable subspaces of the model. Instable space on short time scale is not the same as for long time evolution. Short time instabilities are usually localized in space, while long time eigenvectors of \rf{eigval}  possesses a global structure.  

The third time scale corresponds to the error growing times more than 100 days. On these scales the model exhibits either non-linear chaotic behavior with exponential growth of all $\lambda(t)$ (as it is the case in the square box), or stable behavior when a perturbation of initial state decreases with time (as it is the case in the Black sea model).

 In order to zoom these time scales, we plot the same data in the Log-Log and  Log-Linear coordinates in \rfg{evt}  on the left and on the right respectively. One can see the error  growth in the square box on this time scale is purely exponential with the same exponent $\lambda(t)=A\exp(0.027t)$. The multiplier $A$ is particular for each parameter, but the exponent is always the same.  This confirms the remark made  in \cite{assimtopo}, \cite{sw-lin}: no matter how the perturbation was introduced into the model, it's long-time growth  is determined by the model's dynamics.

Comparing the evolution of the sensitivity of the model to different parameters, we see that on small scales it is the bottom topography that the model is the most sensitive to (thin solid line in  \rfg{evt}). An error in the topography produces 13 times bigger perturbation in the model state than a similar error in the model's initial conditions (thick solid line in  \rfg{evt}). However, $\lambda(t)$ does not grow at all on medium scales due to significant changes in the eigenvectors pattern. This leads to  the fact that on long scales, the sensitivity of the model to the bottom topography is about 2 times lower than the sensitivity to initial conditions. 

On the other hand, the sensitivity of the model to the discretization of operators near the  boundary exhibits the opposite behavior. On short scales, corresponding $\lambda$ is 2 times lower than    $\lambda$ obtained for perturbations of  $\phi_0$, but there is no stagnation of the growth on the middle scales. As a result, we see that the model is 4 times more sensitive to $\alpha$ than to $\phi_0$ for long error growing times.  Moreover, small perturbation of initial conditions decreases in the Black sea configuration, while the perturbation of $\alpha$  results in an increasing perturbation of the solution. 

\section{Modification of the boundary conditions}

As it has been noted, it is useless to analyze  the set of obtained coefficients $\alpha$   to understand  the modification of the boundary conditions. Instead of this,  we shall see the difference between velocity fields with classical and with optimal coefficients $\alpha$ similarly to  \cite{sw-nl}. This difference has been averaged in time over 200 days time interval in order to reveal persistent modifications of the flow produced by the optimal discretization. 
 
This average difference of the velocity together with the original velocity are presented in \rfg{diff-vel}. We zoom  the Southern part of the Black sea because it is in this region  the difference shows the biggest values reaching 5 $\fr{cm}{s}$ while in the middle of the sea it rarely exceeds 1  $\fr{cm}{s}$. 

\begin{figure}[h]
  \begin{center}
  \begin{minipage}[r]{0.48\textwidth} 
  \centerline{\includegraphics[angle=-90,width=0.95\textwidth]{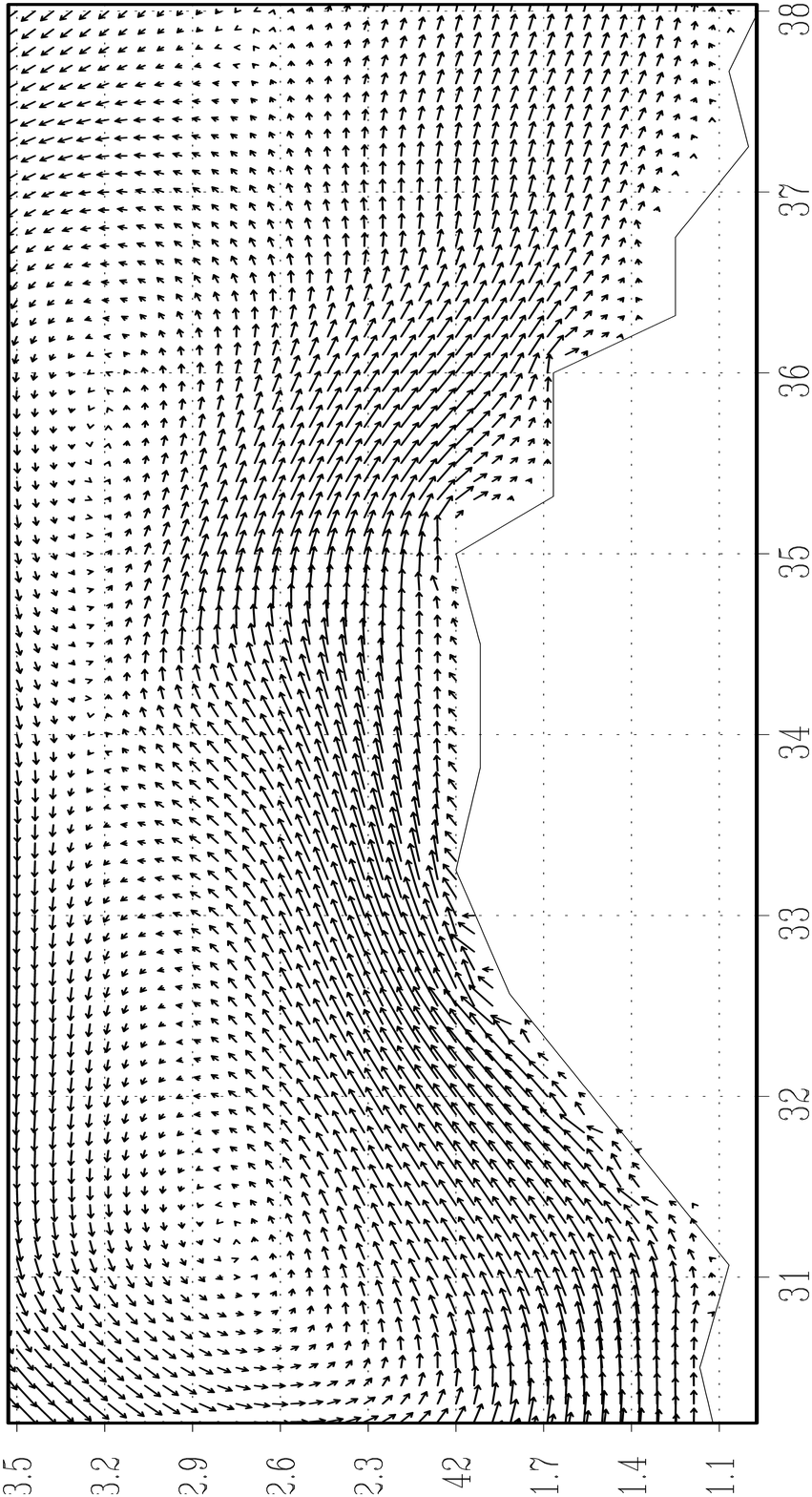}}
  \end{minipage} 
   \begin{minipage}[r]{0.48\textwidth} 
  \centerline{\includegraphics[angle=-90,width=0.95\textwidth]{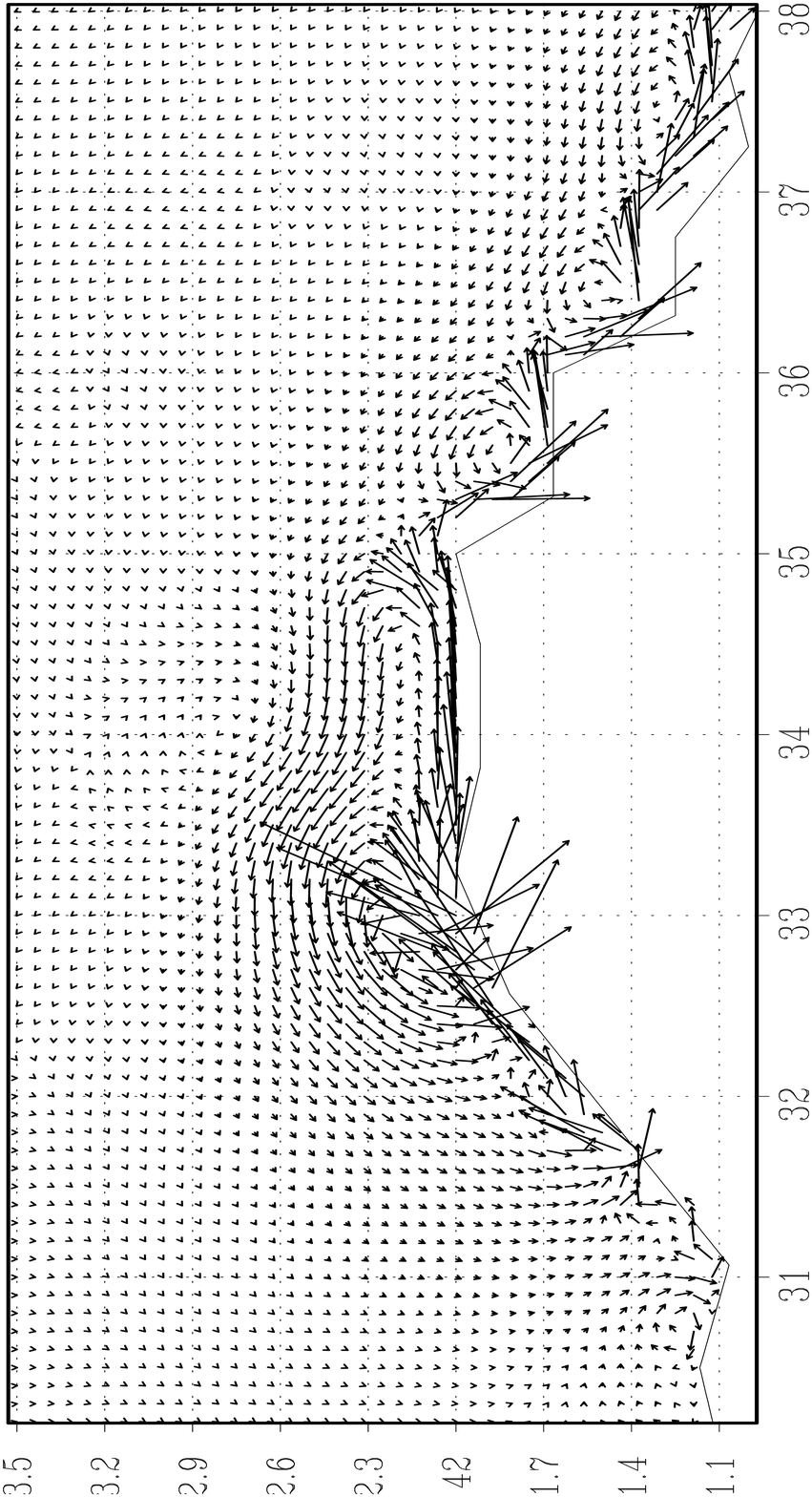}}
  \end{minipage} 
  \end{center} 
  \caption{Original velocity field (left) and its  modification (right)  near the boundary   }  
\label{diff-vel}
\end{figure}

We can note several principal features of the flow that have been modified by   boundary conditions. First, we can see a strong  current on the boundary. The slip condition (vanishing tangential velocity) has been replaced by a permanent current along the boundary. Moreover, impermeability condition has also been modified. The flow is now  allowed to leave  the domain ensuring, however, the global mass balance.   One can see a strong persistent vortex centered at 
$42.2^\circ$N, $32.8^\circ$E which southern part crosses the boundary resulting in not only tangential but normal flux also.    Similar vortices with lower amplitude can also be seen in places where the boundary changes direction. Optimal discretization allows the flow to cross the boundary in places where the direction change is not smooth. 

Tangential velocity component is amplified in the direct vicinity of the boundary. In these nodes we see a strong eastward flow that was  forbidden by the boundary conditions in the classical formulation of the model. On the other hand, the eastward velocity is lower at nodes distanced by several grid cells from the boundary. At these nodes we see westward flow in the difference of  the optimally discretized and classical models. That means the flow is moved towards the boundary, allowing more optimal  representation of a thin current on a coarse grid that brings the  model solution towards observations.

\section{Conclusion}

The comparative study presented in this paper shows the  influence of different model parameters on the solution. The study is confined to the analysis of a low resolution model with a rather limited physics.  Consequently we must acknowledge the results may be valid only in the described case. Additional physical processes (baroclinic  dynamics,  variable density due to heat and salinity fluxes, etc.) may modify  results of this study revealing other parameters the model may be sensitive to.   

The main conclusion we can made from  this comparison is the important role played by the boundary conditions on rigid boundaries. Almost all experiments show that the model is the most flexible with respect to control of coefficients $\alpha$, this control allows us to bring the model's solution  closer to the solution of the high-resolution model or to the observed data.   

Optimal $\alpha$ found in the assimilation window remain optimal long time after the end of assimilation improving the forecasting ability of the model. We could see that the fourth order model in the square box allows us to divide by two  the forecast error of the  20 days forecast. Optimal $\alpha$ obtained in one month assimilation remains optimal even for a one year run of the Black sea model. 

  Finally, the  long time sensitivity of the model's solution to $\alpha$ exceeds the sensitivity to almost all other parameters including the sensitivity to initial conditions. A perturbation of $\alpha$ of a given small norm results in a bigger perturbation of the model's solution than a perturbation of some other parameter of an equal norm. 
 
   However, we could see that the influence of boundary conditions  is only important on long time scales, i.e. time scales that exceeds the characteristic time of the domain. In both experiments presented above the characteristic  time was approximately equal to 5 days and in both experiments the sensitivity to $\alpha$ becomes important on scale longer than 5 days. On the other hand, on short scales,  it is the bottom topography that influences the most the model's solution. Both in the Black sea and in the square box the sensitivity to topography is approximately 40 times more important than the sensitivity to $\alpha$. 
   
In addition to that, we should note that usually prescribed boundary conditions (impermeability and no-slip conditions have been used here as the initial guess for the minimization of the cost function) seem not to be optimal for    the model. As we can see in \rfg{valj}, modifying $\alpha$ we can bring the model much closer to the high resolution model or to the observational data. But, the numerical scheme is strongly  modified in the assimilation process violating even impermeability condition.

Taking into account an important influence of the numerical scheme that introduces boundary conditions into the model,  it is reasonable to think about identification of the  optimal scheme by data assimilation process instead of prescribing classical boundary conditions. 
  
{\bf Acknowledgments. } Author thanks  Gennady Korotaev from   Marine Hydrophysical Institute, National Academy of Sciences of
Ukraine for providing the model parameters and data for the upper layer model of the Black Sea.


\end{document}